\documentclass[a4,12pt]{article}

\usepackage{amssymb}

\usepackage{xcolor} 
\usepackage{amsmath}
\usepackage{subfigure}
\usepackage{hyperref}

\usepackage{graphicx}
\graphicspath{{fig-bactCellSizeDistr/}}

\title{Effect of Replication Fork Dynamics on {\it Escherichia coli} Cell Size} 

\author{Chaitanya A. Athale,\\Div. of Biology, IISER Pune,
              Dr. Homi Bhabha Road, Pashan,\\
              Pune 411008, India.\\
              Tel.: +91-20-2590-8050,
              Fax: +91-20-2586-5086,\\
              Email: cathale@iiserpune.ac.in}
\date{}

\begin{document}

\maketitle


\section{Abstract}
The variability in cell size of an isogenic population of {\it Escherichia coli} has been widely reported in experiment. The probability density function (PDF) of cell lengths has been variously described by exponential and lognormal functions. While temperature, population density and growth rate have all been shown to affect {\it E. coli} cell size distributions, and recent models have validated a link between growth rate and cell size through DNA replication, cell size variability is thought to emerge from growth rate variability. A mechanistic link that could distinguish the source of stochasticity, could improve our understanding of cell size regulation. 

Here, we have developed a population dynamics model of individual cell division based on the BCD, birth, chromosome replication and division model, with DNA replication based on the Cooper and Helmstetter (CH) multi-fork replication. In our model, stochasticity in the model arises solely from the dynamics of DNA replication forks. We model the forks as two-state systems- stalled and recovered. Our model predicts an increase in cell size variability with growth rate, consistent with previous experimental reports. 
We perturb the model to test the effect of increased replication fork (RF) stalling frequency, or uncoupling RF stalling from the cell-division machinery. Indeed, despite ignoring DNA and protein segregation asymmetry, the model can faithfully reproduce quantitative changes in cell size distributions. In our model, multi-fork replication produces multiplicative `noise' and provides a mechanism linking growth rate and cell size variability. 

{\bf Keywords:} cell size; bacterial cell cycle model; population variability; BCD mode; replication stalling; stochastic replication 



\section{Introduction}
\label{intro}
Size and shape cells are characteristic for a given cell type and is a complex phenotype, depending on multiple genetic and non-genetic factors. The determination of cell size in the single-celled bacterium {\it Escherichia coli} by genetic, metabolic and environmental factors has been studied in great detail and reviewed extensively \cite{Chien:2012aa}. Most studies have focussed on the high degree of precision homeostasis observed in bacterial cell size, driven by the {\it minCDE} system, that has been shown to be a critical player in such accurate division site placement. In the absence of MinCDE, mutant {\it E. coli} cells were found to form FtsZ rings as a prelude to division, wherever the nucleoid was absent \cite{Yu:1999aa}. Indeed, analysis of variations in cell septum formation as a function of cell length suggest a high precision in cell center finding during equational division of 2.9\% of the cell length  \cite{Guberman2008}. This was also shown to be robust to deformation of cell shape, with a narrow the daughter cells showing exactly half the ratio of the mother cell prior to division with very narrow distribution \cite{Mannik:2012aa}. All these results however focussed on the convergence of cell sizes to a mean value observed in experiment.

The observed heterogeneity of short and long cells in a population, with occasional filamentous or extremely long cells \cite{Koch1966,Koppes1978}, is however considered to be a complex result of variation in cell physiology and the environment. 

\begin{figure}[ht!]
\begin{center}
	\subfigure{\includegraphics[width=0.45\textwidth]{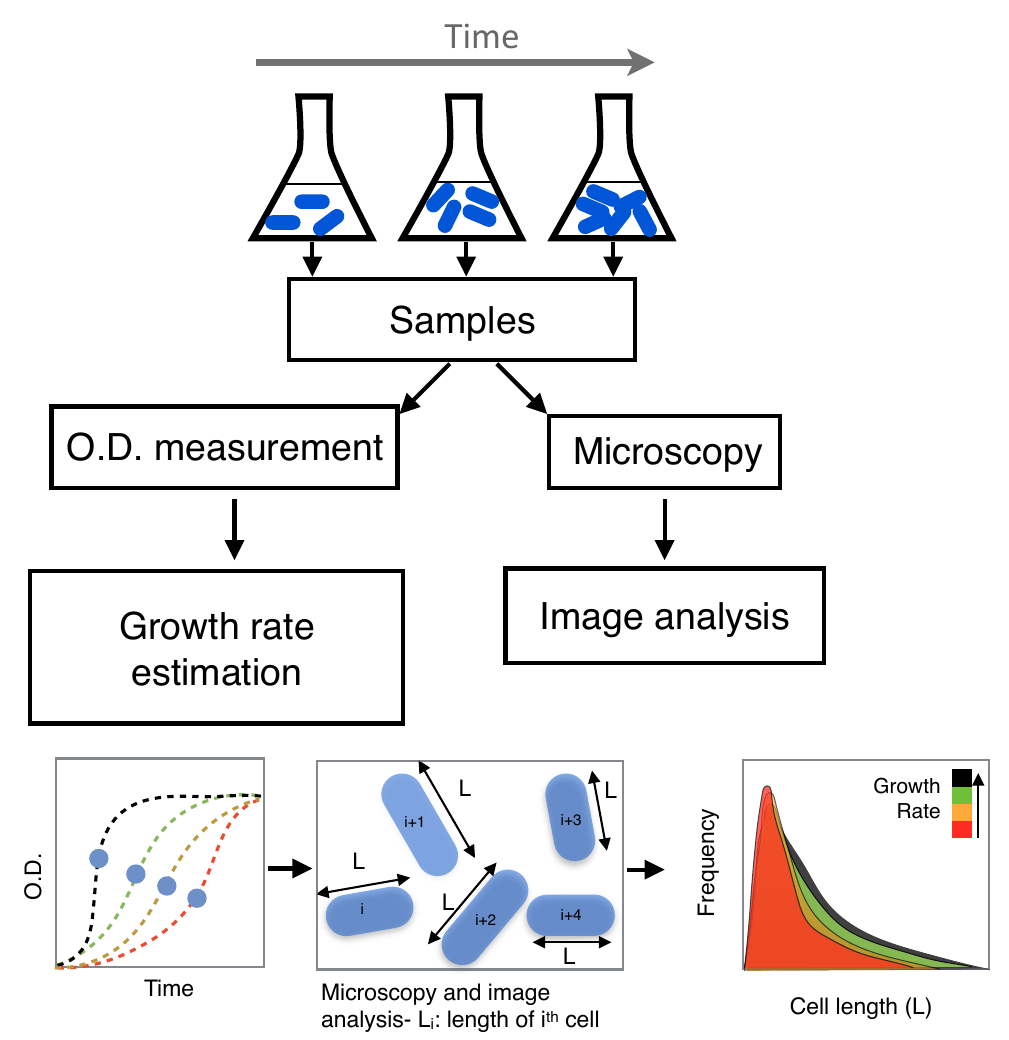}  \label{fig:expSchem} }
	\subfigure[]{\includegraphics[width=0.5\textwidth]{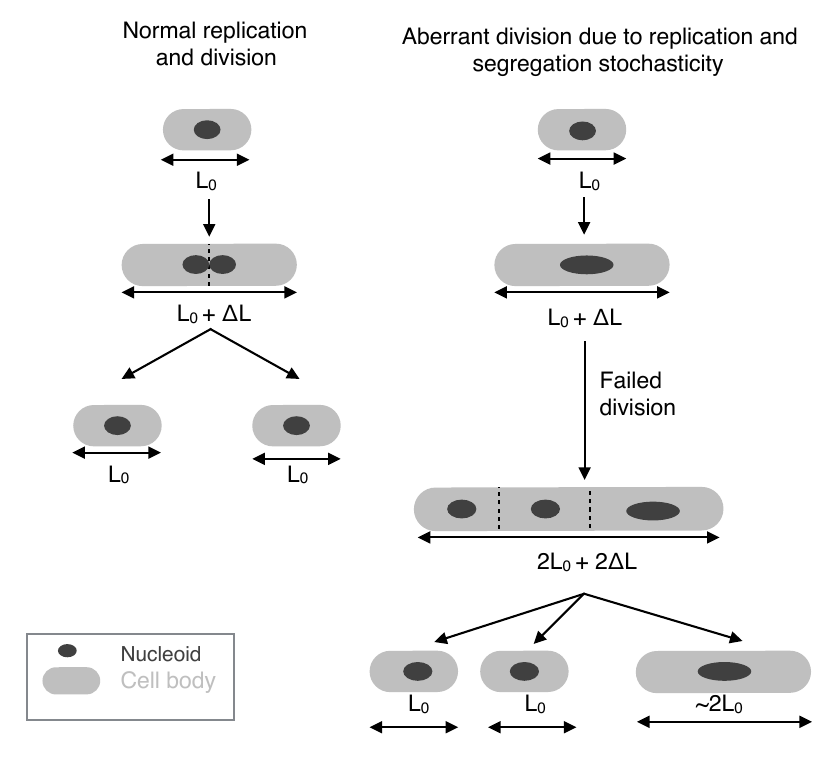} \label{fig:aberrDivision}}
\caption{{\it E. coli} cell size variability. \subref{fig:expSchem} Experiments to cell length variability are based on monitoring population density and growth rates of liquid batch cultures of {\it E. coli} and sampling population cell lengths by microscopy and image analysis. In previous work we have shown higher growth rates result in more variable cell length distributions \cite{Athale2011,Gangan:2017aa}. \subref{fig:aberrDivision} The schematic illustrates the potential for emergence of cell length variability during the replication and division process, where a newborn cell length $L_0$ increases to $\sim 2 L_0$ just prior to division. We argue for a model in which deterministic DNA replication will results in two cells of the same birth length ($L_0$) due to the precision of FtsZ division site placement by the MinCDE system. However, stochastic replication or segregation defects can failed division events, that result in cells of varied length $\geq 2 L_0$. $\Delta L$ is the constant increment in cell length per based on the `incremental' model \cite{Amir2014,Taheri-Araghi2014}.}
\label{fig:exptSchem}       
\end{center}
\end{figure}

\subsection{{\it E. coli} cell size control}
Cells of the model bacterium {\it Escherichia coli} have described as sphero-cylinders with a length of 2 $\mu m$ and width of 1 $\mu m$ at birth. A single newborn {\it E. coli} cell during growth elongates to twice its length and divides symmetrically into two daughter cells. Cells sampled at the same growth stage (e.g. birth) are variable primarily in cell length (L), while the width appears to be constant as seen in measurements made in light- and electron-microscopy \cite{Cullum1978,Grover1987,Kaya2009}. 

\subsection{Effects of nutrient availability and population density on cell size}
Environmental factors such as low bacterial density \cite{Maclean1961} or a shift to richer media \cite{Kubitschek1990} have been shown to increase the proportion of long-cells. The average cell mass and DNA content is independent of temperature ($25^oC$ and $37^oC$), but a change in the growth medium can result in change in mass and DNA content as seen in {\it Salmonella typhimurium} \cite{Schaechter1958}  and {\it E. coli} \cite{Shehata1975}. On the other hand, a study on the effect of temperature on {\it E. coli} sizes showed cells at $22^oC$ are shorter than at $37^oC$ \cite{Trueba1982}. At the same time, growth rate alone has been shown to correlate with increased cell size and multiple nucleoids in {\it Salmonella} \cite{Schaechter1958}. These observations with respect to the effect of temperature and nutrients on cell size imply gaps in both measurement and our theoretical understanding. Recent improvements in light microscopy now allow visualizing sub-cellular processes and their dynamics even in cells as small as {\it E. coli} \cite{Wang2008}. Coupled with computational image-analysis methods \cite{Eils2003}, a systematic quantification could result in a disambiguation of these results.

\subsection{Molecular basis of {\it E. coli} cell size determination}
The molecular mechanisms by which nutrient availability determines bacterial cell size has recently been shown to be mediated in {\it E. coli} by phosphoglucomutase (pgm) \cite{Hill2012} and glucosyltransferase OpgH \cite{Hill2013} and in {\it Bacillus subtilis} by the protein UgtP \cite{Weart2007}. While these factors are known to affect the central tendency of cell size, they do not explain effects that change the spread of the distributions. Additionally most pathways regulating {\it E. coli} cell size, including nutrient sensors, converge on FtsZ as the primary molecular determinant of newborn cell size \cite{Lutkenhaus1980,Bi1990,Lutkenhaus2007}. The regulation of division in {\it E. coli} by oscillations of MinCDE proteins \cite{Raskin1999,Varma2008} and nucleoid occlusion by the SlmA-DNA complex \cite{Woldringh1990,Bernhardt2005} is well established. Additionally an indirect effect has been observed based on SulA inhibition of FtsZ, triggered by RecA dependent SOS response \cite{Huisman1984}. While elongated cells (length $>8$ $\mu m$) in wild-type populations have been reported, recent quantification of the cell-division septum has estimated a deviation from the mid-cell by 10\% of cell length \cite{Guberman2008}. Theoretical modeling could serve to reconcile the relative importance of these mechanisms and lead to a deeper understanding of the regulation of a complex phenotype such as cell size.

\subsection{Models of bacterial cell size regulation}
Mathematical models of bacterial cell size can be broadly thought of as those that address the mean size of a cell and those that address the distribution in a population. Cell mass distributions in growing populations of {\it Bacillus cereus} were proposed by Collins and Richmond, where the distribution of cell masses (m) is defined as $\theta(m) \propto 1/m^2$ to result in stable cell length distributions \cite{Collins1962}. A model of DNA replication during rapid growth of {\it E. coli} has been described first by Cooper and Helmstetter's (CH) to require multi-fork replication, i.e. more than one copy of the single origin of replication is progressing in cells, to compensate for the maximal rate of DNA replication within the generation time \cite{Cooper1968}. A computational model that illustrated this replication process based on deterministic modeling demonstrated the exact counts expected from an arbitrary growth rate \cite{zaritsky2011}, but this was not coupled to cell division and size. The frequency distribution of cell lengths typically shows a positive skew due to the presence of long cells which have been defined in literature as having $L > 8$ $\mu m$ \cite{Cullum1978}. Models that attempted to explain this distribution of cell sizes either assumed asymmetric cell division \cite{Cullum1978} or stochastic partitioning of molecular components during cell division \cite{Huh2010}. With recent developments in microfluidics, the measurement of single-cell bacterial growth kinetics \cite{Wang2010} have been used to distinguish between the `timer' and `sizer' mechanisms that had been previously proposed to govern bacterial cell sizes \cite{Robert2014}. Alternatively, an analytical model which invoked volume change as a governing criterion for cell size determination, and referred to as an `incremental model' or `adder' model, has been shown to match cell size and its dynamics \cite{Amir2014,Taheri-Araghi2014}. The accumulation of an initiator per replication origin, has been hypothesized to form the basis of the volume increment and match growth rate dependence of cell size \cite{Ho:2015aa}. More recently, a model of DNA replication initiation initiated by a volume sensing mechanism has explained the apparent correlation of the `adder' model with fast growth, while also explaining deviations from `adder' during slow growth \cite{Wallden:2016aa}. In this model Wallden et al. invoke the role of multi-fork replication and show that newborn cells add a volume that is based on growth rate sensing. Through this model they indeed recover growth rate dependence of cell size variability, which they attribute to a variability in growth rates. Due to the mismatch between this otherwise successful model of Wallden et al., a novel model has been proposed of concurrent and competing division and replication cycles that govern cell size distributions \cite{Micali:2018aa}. While indeed cell-cell variation in growth rates is expected, from the phenomenological models, it is not clear how these lead to cell size variability. Given our previous work demonstrating the increased percentage of long cells, when cells are treated with hydroxyurea, increasing replication stochasticity \cite{Gangan:2017aa}, this mechanism could help improve our understanding of the mechanism that links growth rate and cell size variability, in addition to intrinsic growth rate variability. 

Therefore, we have developed a stochastic model of {\it E. coli} cell division in a growing population of simulated cells. The model spans the scales of sub-cellular replication fork dynamics, whole cell scale birth, growth and division dynamics and population growth of cells in a limited nutrient pool, with the only source of variability based in replication processivity or `stalling'. Experimental measurements of cell size distributions of {\it E. coli} mutants of the SOS response pathway, are used to test the model predictions.

\section{Theory: Multi-scale model of bacterial population growth}
\label{sec:theory}
A multi-scale model of a population of individual cells or `agents' growing in length and dividing has been developed. The instantaneous growth rate (doubling time) is modeled as a variable, based on the logistic model, which depends on the size of the population scaled by the available nutrients. This reproduces the growth phases of a population growing in the presence of limited nutrients seen in typical batch cultures- lag, log and stationary. A modified birth, chromosome-replication and division (BCD) model of the bacterial cell cycle has been developed, in which chromosome-replication has been explicitly modeled by the discrete, stochastic dynamics of replication forks (RFs). RF dynamics is coupled to cell division through a model of cell division inhibition, based on the role of RecA-SulA inhibition of FtsZ. Model parameters are taken from literature where available, and the remaining are optimized (Table \ref{tab:simparam}).

\begin{figure}[ht!]
\begin{center}
 	 \subfigure[]{    \includegraphics[width=0.3\textwidth]{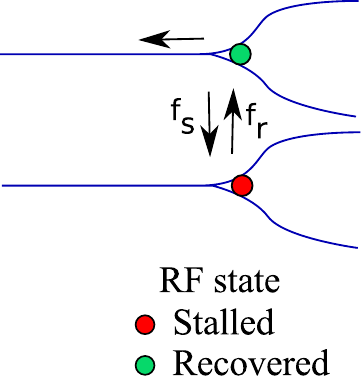} \label{fig:rf} }
 	 \subfigure[]{    \includegraphics[width=0.3\textwidth]{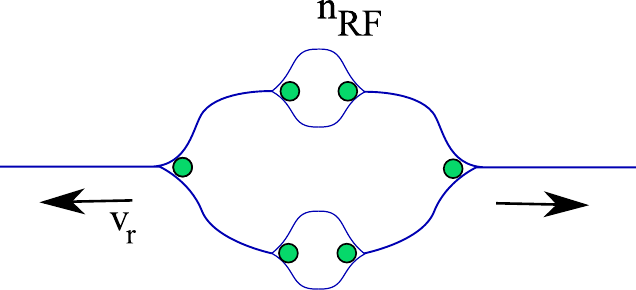} \label{fig:multifork} }\\
   	 \subfigure[]{    \includegraphics[width=0.3\textwidth]{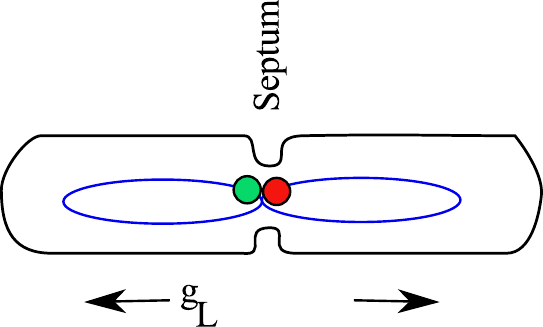} \label{fig:divgro} }
	 \subfigure[]{    \includegraphics[width=0.3\textwidth]{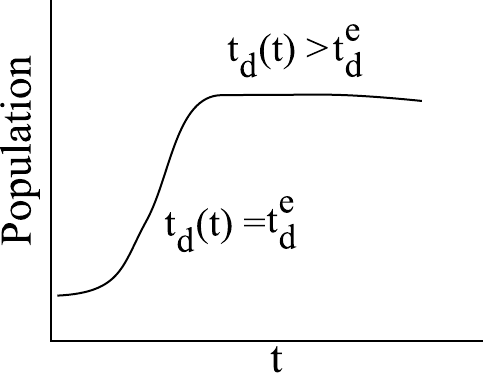} \label{fig:popgro} }
\caption{Model of replication, division and population growth of {\it E. coli}. \subref{fig:rf} DNA replication can proceed if the replication fork (RF) is in the recovered state (green) but not if it is in the stalled (red) state. The transition frequency of forks to a stalled state is determined by $f_s$, the frequency of stalling, and the transition to recovered state by $f_r$, the frequency of recovery.  \subref{fig:multifork} Replication proceeds at a speed given by $v_r$ bidirectionally on the genome. \subref{fig:divgro} Cell growth is governed by the growth rate ($g_L$) and the formation of a septum results in cell division. The cell can only divide if all RFs are in recovered (green) state. \subref{fig:popgro} The population size (N) as a function of time (t) follows the logistic model due to a variable doubling time ($t_d(t)$) based on Equation \ref{eq:pop} in this schematic plot. The growth curve enters the `stationary phase' when $t_d(t)$ exceeds $t_d^e$ due to the population size ($N(t)$).}
\label{fig:modelschem}       
\end{center}
\end{figure}

\subsection{Stochastic multi-fork replication}
\label{ssec:replication}
DNA replication in {\it E. coli} is modeled by a two-state dynamics of an individual RF. An RF is either in a stalled or replicating state (Figure \ref{fig:rf}). In the replicating state, DNA is replicated by the RF with a genome replication speed ($v_r$). The value of $v_r$ is taken from the {\it E. coli} DNA PolIII holoenzyme replication speed \cite{ODonnell1990,Pham2013}. The frequency of stalling ($f_s$) determines the transition of an RF from replicating to stalled state and the frequency of recovery ($f_r$) determines the transition back from a stalled to a replicating state. The values of $f_s$ and $f_r$ were optimized to qualitatively reproduce experimental cell length ranges in wild-type cells. These frequencies are compared at each iteration step to a random number drawn from $U[0, 1]$ to determine the state of the RF. 

During rapid growth, {\it E. coli} is known to initiate more than one pair of RFs (Figure \ref{fig:multifork}). The number of RFs ($n_{RF}$) at any given time in a cell is modeled by modifying Cooper and Helmstetter's (CH) model of multi-fork replication \cite{Cooper1968} as:

\begin{equation}
    n_{RF} =
    \begin{cases}
	2 & \text{if } t_{d}(t) \geq C + D\\
	2\cdot \lceil{1+(C+D)/t_{d}(t)} \rceil & \text{if} t_{d}(t) < C + D
     \end{cases}
     \label{eq:rfmulti}
\end{equation}

where C and D is the time taken for chromosome replication and cell division respectively and $t_{d}(t)$ is the instantaneous doubling time. The value of $n_{RF}$ is mapped to the largest following integer, with the factor 2 ensuring it is an even number, since replication forks are always known to be initiated in pairs, based on Cooper \& Helmstetter. The values of the C- (40 minutes) and D-periods (20 minutes) are set as constants based on previous measurements in fast growing cultures \cite{Michelsen2003}.


\subsection{Cell elongation model}
\label{ssec:cellgro}
The growth in length of {\it E. coli} cells measured in experiment has fits bi- and tri-linear rates ranging  between 0.1 and 0.2 $\mu m/min$ \cite{Reshes2008a}. These rates are constant for cells in a given condition. We model the growth in cell length ($g_L^i$) of the $i^{th}$ cell with a single linear growth rate, which averages over the reported bi- and tri-linear fits. Since our model assumes batch culture with limited carrying capacity, while $g_L^i$ is constant for a single cell, this value is different for cells depending on nutrient availability at the time of birth. We assume $g_L$ decreases with increasing population size, similar to the effect on doubling time, and hence our model of $g_{L}^i$ depends on the instantaneous doubling time $t_{d}^i(t=birth)$, at the time of birth of the $i^{th}$ cell and it's length at birth ($L_b^i$) as follows:

\begin{equation}
	g_L^i = L_{b}^i / t_d^i(t=birth)
	\label{eq:elongR}
\end{equation}

Thus, for a single cell $g_L^i$ is constant in a manner consistent with the ``incremental model'' of bacterial cell volume growth \cite{Amir2014}. The doubling time of a single cell is taken from the population doubling time ($t_d(t)$) of a batch-culture. The value of the doubling-time in our model is variable, increasing with increasing population size (i.e. slower growth), due to the reduced per cell nutrients availability. The $\sim1/t_d$ dependence of $g_L$ is based on the widely observed effect of reduced nutrients to reduce single cell wall-growth rates, due to the coupled nature of the pathways that regulate cell wall synthesis and nutrient sensing and metabolism \cite{Wang2009,Yao2012}. In contrast, the ``incremental model''  makes the assumption that the nutrient is not limiting, resulting in a continuously growing culture. At a given time point $t$, the $t_d^i$ is uniform for the population i.e. doubling time does not contribute to the variability in cell elongation rates. However, $L_b^i$ may differ between cells, due to the replication-division coupling driven by replication stochasticity. Thus, in our model the cell elongation rate is constant for each cell, and variations between cells are the result of aberrant replication or growth phase of the population.

\subsection{Cell division model}
\label{ssec:celldiv}
The cell in our model divides based on a decision represented by $\Gamma$; while $\Gamma = 0$, they continue to grow, and if $\Gamma=1$ cells divide. Three primary mechanisms known to regulate division onset are used to determine the state of $\Gamma$: (i) cell age ($t_{cell}$), (ii) the length of the genomic DNA that has been replicated in the cell (${DNA}_{cell}$) and (iii) the state of the replication fork state (stalled/recovered). These conditions are modeled as follows:

\begin{equation}
  \Gamma  = \left\{ 
   \begin{array}{rl}
         	 	1 &  \text{   if} \left\{ 
		 \begin{array}{ll}
		 		t_{cell}(t) \geq t_d^i \\ 
		          {DNA}_{cell}(t) = 2\cdot {DNA}_{Ecoli} \\
		         \min_{1}^{n} [state(RF)] =1\\
		        \end{array} \right. \\
			0 & \\   
	     \end{array} \right.
\end{equation}

where $t_{d}(t)$ is the instantaneous doubling time of cells, ${DNA}_{Ecoli}$ is the complete genome length of {\it E. coli} and $\min_{1}^{n} [state(RF)]$ finds the minimal value of the state (0: stalled, 1: recovered) of all replication forks, where $n$ is the total number of replication forks. The first condition is based on the BCD cell cycle \cite{Cooper1968}, where division occurs if the cell age corresponds to or exceeds the doubling time at birth of that cell ($t_d^i$). The second condition is based on `nucleoid occlusion' \cite{Bernhardt2005}, which prevents cell division until one round of replication is completed. This condition is imposed to only allow division if the oldest replication fork pair in the cell.. and the RF state models the inhibition of division by the SOS-pathway. In the model, when $\Gamma=1$ cell division proceeds instantaneously, since we assume rapid kinetics of FtsZ polymerization (Figure \ref{fig:divgro}). In this way, our cell division model couples the stochastic DNA replication with the deterministic parts of the model. In the model, cell length variation beyond the expected range (2 to 4 $\mu m$) arises from a failure of cells to divide. Such an elongated cell may divide at the next round of replication if the conditions are satisfied. The model does not impose a `sizer' or `timer' mechanism, instead a cell size `emerges' from the cell cycle model.

\subsection{Population dynamics model}
The growth of a population of discrete cells with finite resource is modeled by a variable doubling time ($t_d(t)$) (Figure \ref{fig:popgro}). Based on the logistic growth model, the doubling time for the whole population at each time point $t$ is calculated as:
\begin{equation}
	t_d(t) = \frac{ t_{d}^{e} }{ ( 1 - N(t)/K ) } 
	\label{eq:pop}
\end{equation} 
The value of the exponential doubling time ($t_d^e$) in the model is constant for a specific environmental condition- temperature or nutrient medium and allows us to test the effect of the maximal growth rate and growth phase on cell length variability. The carrying capacity ($K$) is fixed for all calculations (Table \ref{tab:simparam}). 

\section{Materials and methods}\label{sec:meth}

\subsection{Data analysis}
Cell length variability was estimated by estimating the sample mean $\mu$ and variance $\sigma^2$ of cell lengths from a growth condition (time, population density). The fano factor (FF) then is $FF=\sigma^2/\mu$. 

\subsection{Simulations}
The simulation was written in MATLAB (Mathworks Inc., Natick, MA, USA) and run on a machine with two 2.93 GHz Quad-Code Intel Xeon processors with 16GB RAM. A single run with $N(t=0)=10$ and $K=1000$ cells respectively and time step $\delta t = 1$ second required 6 hrs to run on an average.

\section{Results}
\label{sec:results}

\begin{figure}[ht!]
\begin{center}
	 \subfigure[]{    \includegraphics[width=0.3\textwidth]{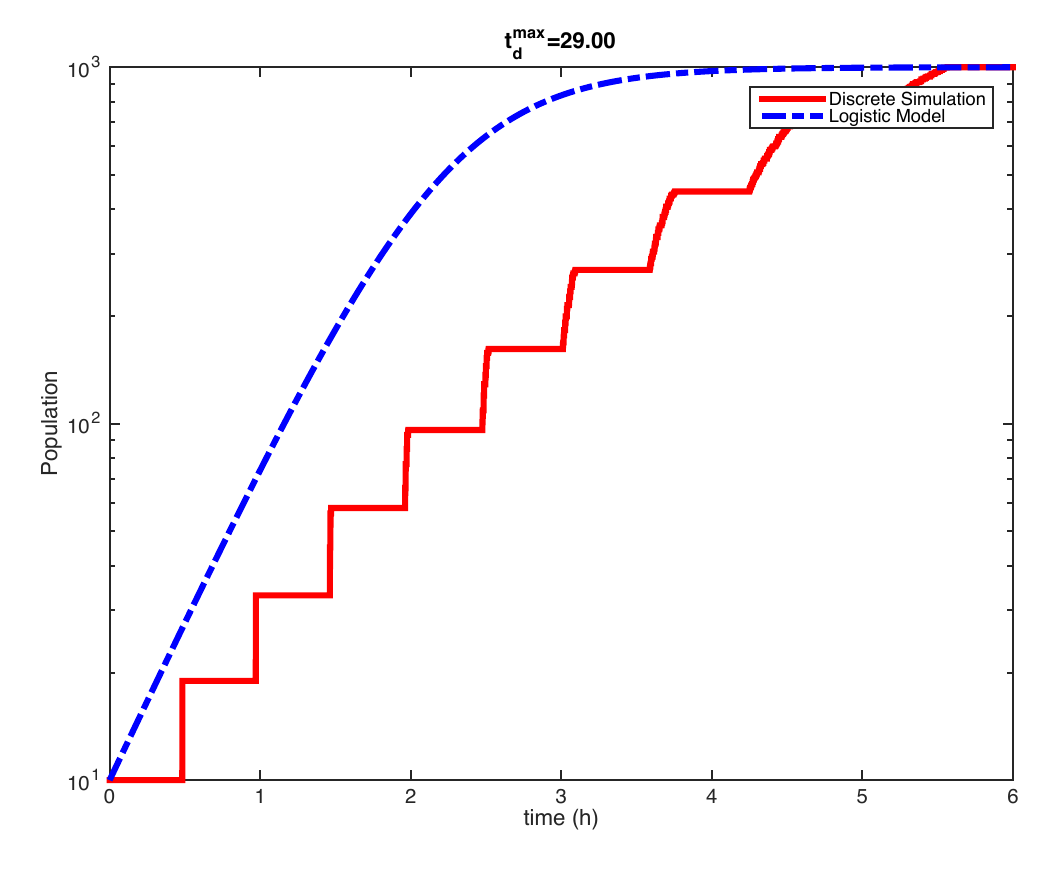}  \label{growthcurve} }
   	\subfigure[]{    \includegraphics[width=0.5\textwidth]{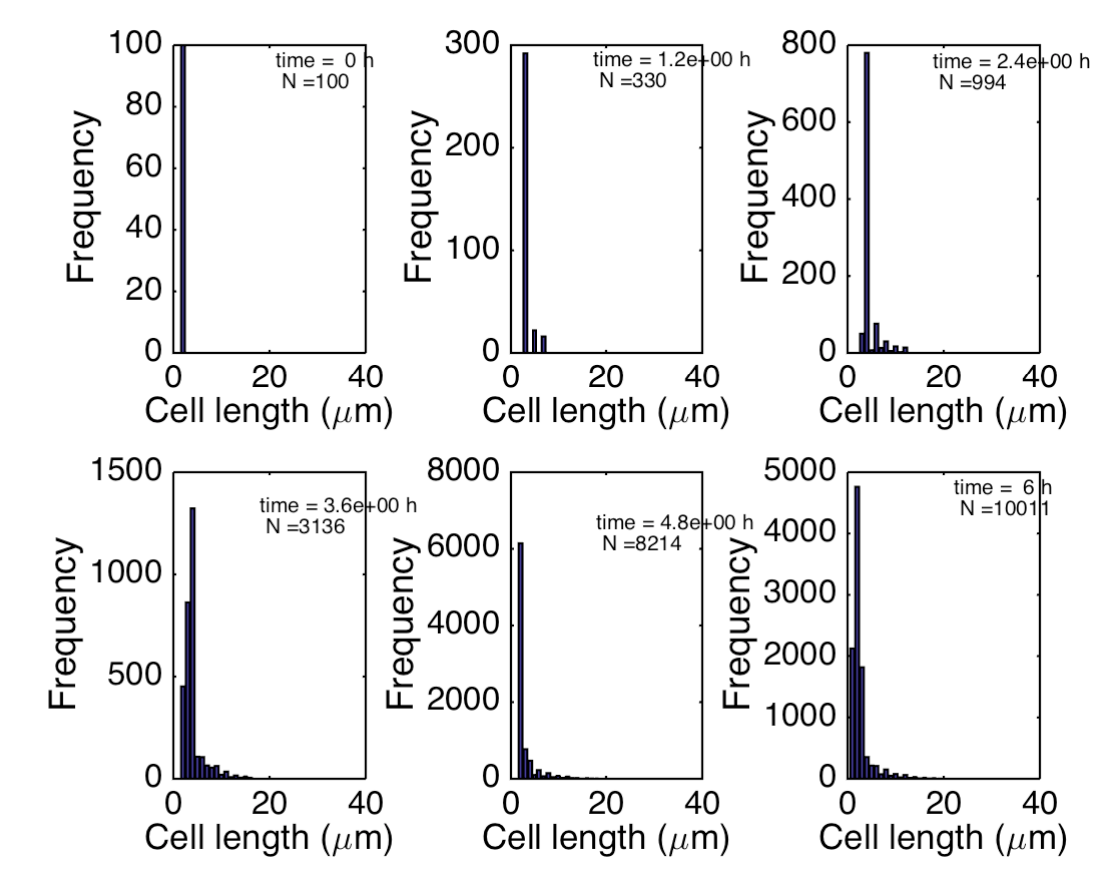}\label{lendistr} }\\ 
	  \subfigure[]{    \includegraphics[width=0.5\textwidth]{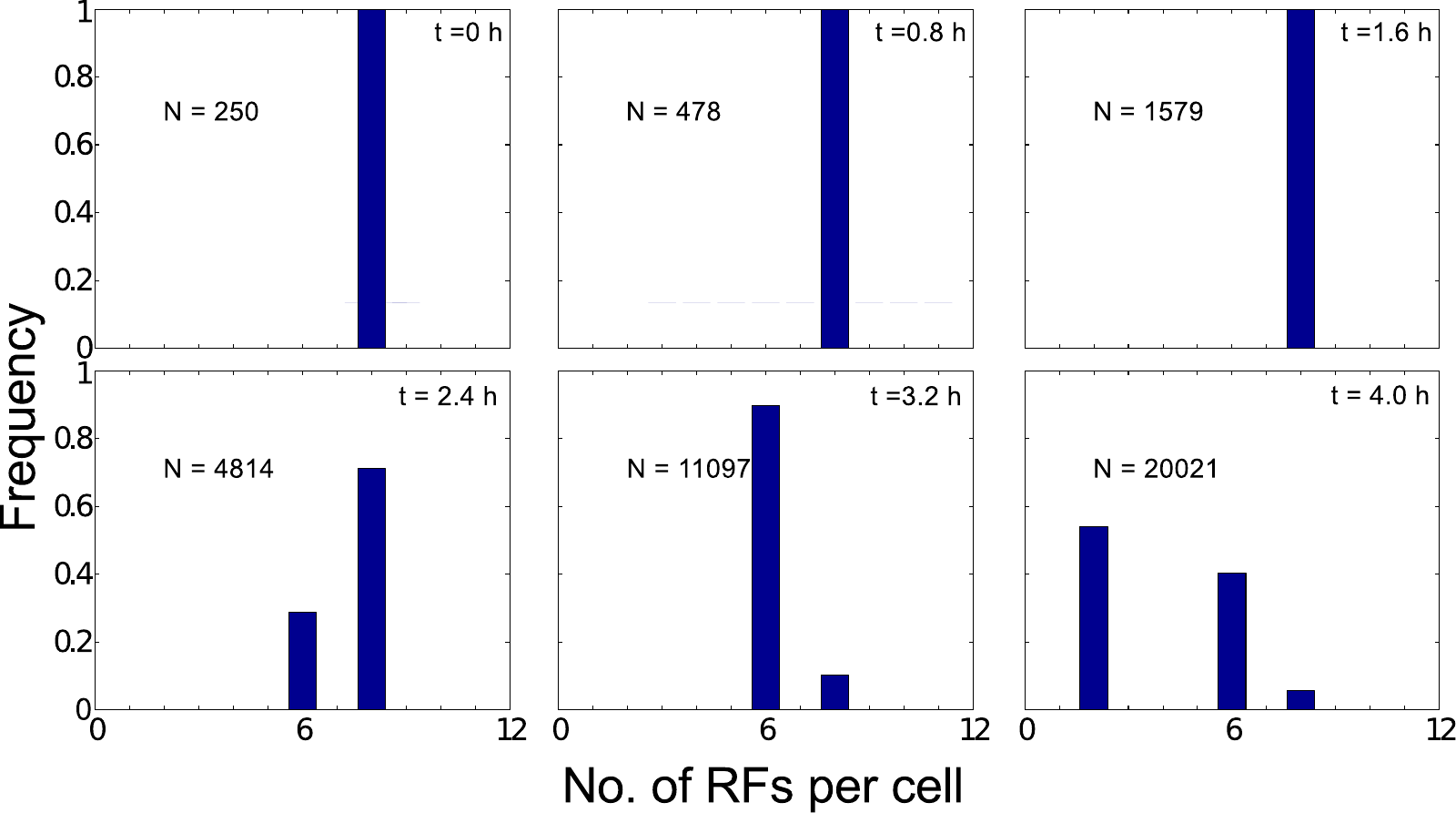} \label{rfFrq} }
	 \subfigure[]{    \includegraphics[width=0.4\textwidth]{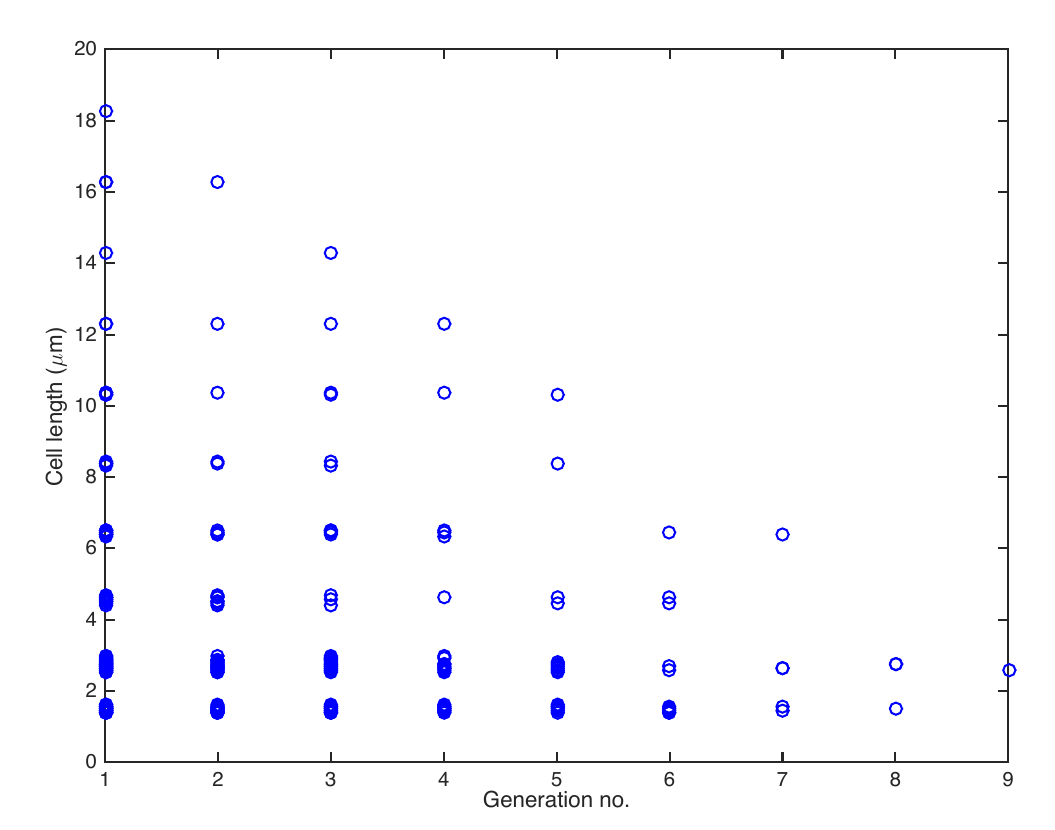} \label{lengthAge}  }
	
\caption{Multiscale simulation of logistic growth, with single cell replication and division dynamics. \subref{growthcurve} The model is simulated with $t_d^e$ 30 min and the population as a function of time from the discrete model (red) is compared to the logistic model (blue) for the same parameters. \subref{rfFrq}-\subref{lendistr} The population frequency distribution of \subref{lendistr} cell lengths and \subref{rfFrq} number of RFs per cell are plotted as snapshots a times $t$. 
       Calculations were performed with simulation parameters based on Table \ref{tab:simparam}.}
\label{fig:simoutput}       
\end{center}
\end{figure}

\subsection{Population cell length distributions in limited nutrient growth}
\label{sec:simcellsize}
In previous experimental work, the effect of population density, growth phase and growth rate had been analyzed using population variability in cell lengths of {\it E. coli} cells grown in batch culture, by simultaneously monitoring the bulk growth by optical density measurements and microscopy combined with image analysis of cells (Figure \ref{fig:exptSchem}). It was shown that cell size variability increased with increasing growth rate and was affected by stochasticity of replication, in a process mediated by RecA and SulA \cite{Gangan:2017aa}. 
To test this model of the effect of replication stochasticity on cell size, we have simulated the multiscale model of cell growth, replication and division. 
Based on the logistic model, the population increases in growth and then saturates as seen in Figure \ref{growthcurve} due to the time dependent change in doubling time ($t_d(t)$) that is inversely proportional to population density at that time ($N(t)$) (Equation\ref{eq:pop}). The single cell elongation rate of the $i^{th}$ cell ($g_L^i$) is inversely dependent on $1/t_d^e$  (Equation \ref{eq:elongR}), and in turn the dependence of In the representative snapshots from growth at $37^o$C ($t_d^e=25$ min) taken every 0.8 hours, in the early phase of rapid growth, each cell has 8 RFs (up-to 1.6 hours) (Figure \ref{rfFrq}). As the population growth slows due to saturation effects, a shift in the distribution of RFs per cell ($n_{RF}=$6 to 8) is seen in 2.4 to 3 h, and near at 4 hours, the majority of cells have $n_{RF}=2$ when growth has reached saturation. Cell sizes of such a population at the start of the simulation are initialized to 2 $\mu m$. As the population grows, the cell lengths begin to be more broadly distributed (Figure \ref{lendistr}). The cell length distributions were fit by a lognormal function, in which the variance increases with time and appears to saturate. The lognormal variance is maximal at 3.2 hours ($\sigma^2=2.37 \mu m^2$) and minimal at 0.8 hours from the snapshots. The multi-scale model thus reproduces the expected population behavior of cell numbers with time at different maximal growth rates. The observed changes in replication fork and cell length distributions suggest a growth phase dependence, which we have analyzed through the growth of the population.

\begin{figure}[ht!]
	\subfigure[]{    \includegraphics[width=0.8\textwidth]{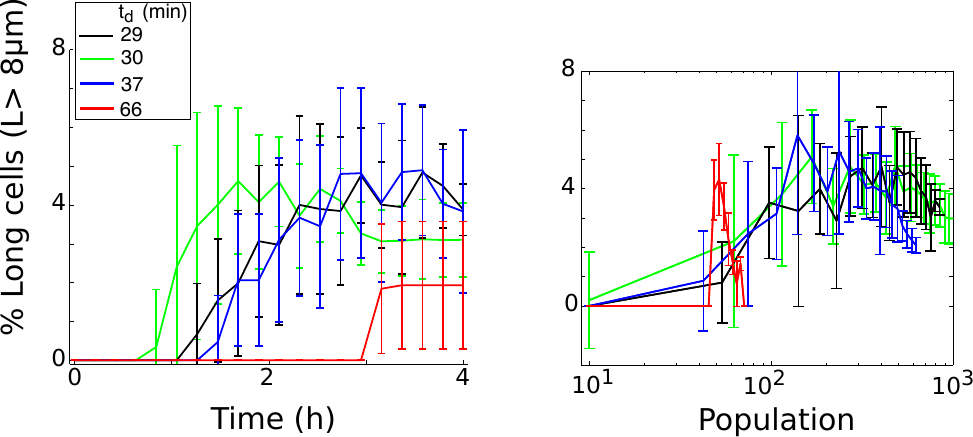} \label{fig:longCells} } \\
   	\subfigure[]{    \includegraphics[width=0.8\textwidth]{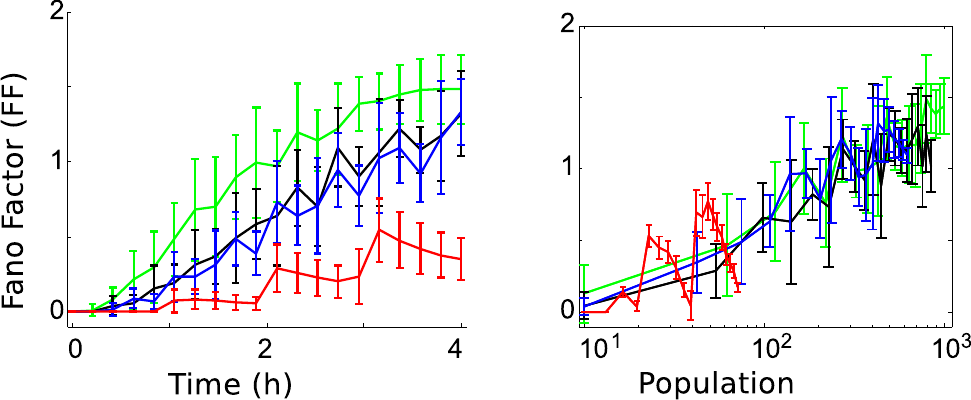} \label{fig:simFano} }
	
\caption{The cell length variability estimated by the Fano Factor ($\sigma^2/\mu$) of cell lengths of the population are plotted as a function of time and population density. The different curves correspond to doubling times, $t_d^{e}$, of  29 min (black), 30 min (green), 37 min (blue) and 66 min (red). The data points are mean $\pm$s.d. (n=10 iterations).}
\label{fig:simNoisetemp}       
\end{figure}
\begin{table}
\caption{Simulation parameters}
\label{tab:simparam}       
\begin{tabular}{p{1cm}p{5cm}p{5cm}p{2cm}}
\hline\noalign{\smallskip}
Symbol & Parameter description & Value & Reference \\
\noalign{\smallskip}\hline\noalign{\smallskip}
K & Carrying capacity & $10^3$ cells & This study \\
$N_i$ & Initial population size & $10$ cells & This study \\
$t_d^e$ & Maximal doubling time & 30, 29, 37 and 66 min & This study \\
$f_s$ & Frequency of RF stalling & 0.005 (w.t.), 0.001 ($\Delta sulA$, $\Delta slmA$) $s^{-1}$ & this study \\
$f_r$ & Frequency of RF recovery & 0.09 $s^{-1}$ (w.t.), 0.05 ($\Delta recA$) & this study \\
$L_g$ & Genome length & $4\cdot 10^{6}$ bps & \cite{Blattner1997} \\
$v_r$ & Genome replication speed & $10^{3}$ bps/s & \cite{ODonnell1990,Pham2013} \\
\noalign{\smallskip}\hline
\end{tabular}
\end{table}

\subsection{Cell size variability with growth phase and rate}
\label{sec:varTemp}
The simulated population cell length distributions appear most variable in the mid-log phase, as seen from the spread in the length frequency distribution snapshots (Figure \ref{lendistr}). On estimating the percentage of elongated cells (cells with length $> 8$ $\mu m$) throughout the entire time-course of the simulation, we find an initial increase, which saturates as a function of time (Figure \ref{fig:longCells}). The mean \%age long cells when $t_d^e$=30 min reaches saturation earlier than $t_d^e$=29 min and followed by $t_d^e$=37 min and 66 min. The \%age long cells of the slowest growing population ($t_d^e$=66 min), does not saturate at all. When the data was plotted as a function of increasing population size, the plots from multiple growth rates appear more comparable to each other, as compared to the time-dependent plots. Similarly, estimating cell length variability by the Fano Factor (FF) as a function of time, shows an increasing trend, in which the different growth rate plots are offset (Figure \ref{fig:simFano}). The time-dependent graphs of both percent long cells and FF as a function population size however collapses all the curves onto the same profile. Surprisingly we find cell length variability for the different maximal growth rates is dominated by the population size, that relates to instantaneous doubling time, and not so much the maximal doubling time ($t_d^e$). Therefore, we predict that experimental measures of cell length variability, when scaled by population should depend on instantaneous growth rates. While this prediction is consistent with published single-cell data, it is based on the independence of DNA replication \cite{Wallden:2016aa}. We therefore proceeded to test the specificity of our model, by perturbing of the parameters of stochasticity of replication, and comparing the resulting cell length distributions with experiments with {\it E. coli}.

\begin{figure}[ht!]
\begin{center}
	  \includegraphics[width=0.7\textwidth]{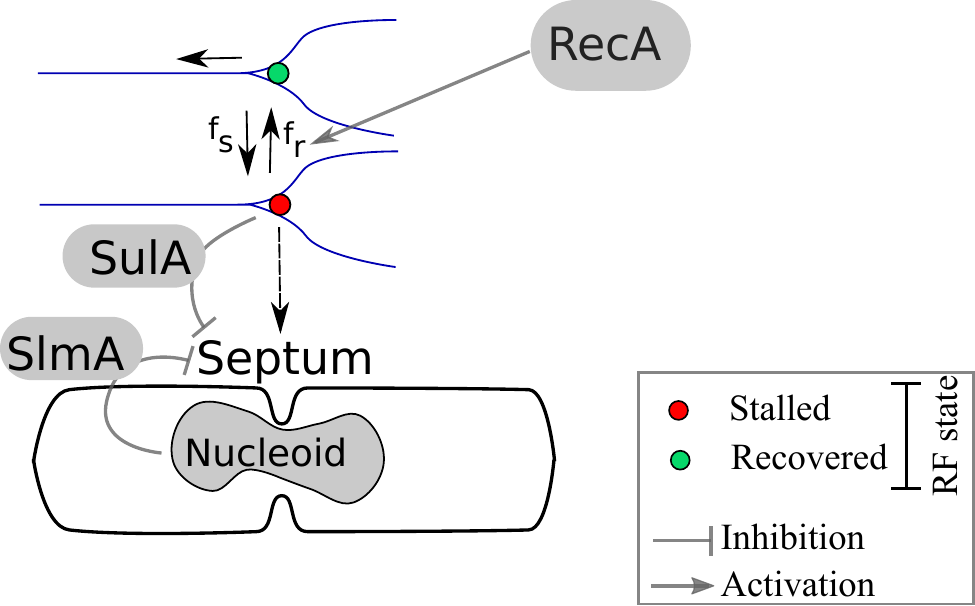}
\end{center}
\caption{{\bf Effect of mutations in dynamics of RF.} The schematic represents the regulation of the two-state model of RF dynamics between stalled (red) and recovered (green) states. Functional {\it RecA} proteins assemble on stalled replication forks and increase the $f_r$, while stalled replication forks result in {\it SulA} mediated inhibition of cell septation by inhibition of FtsZ assembly. {\it SlmA} acts to inhibit cell division by FtsZ driven septum formation if the nucleoid has not completely segregated after a round of DNA replication.}
 \label{fig:recsulModel}        
\end{figure}
	 
\begin{figure}[ht!]
 	 \subfigure[]{    \includegraphics[width=0.5\textwidth]{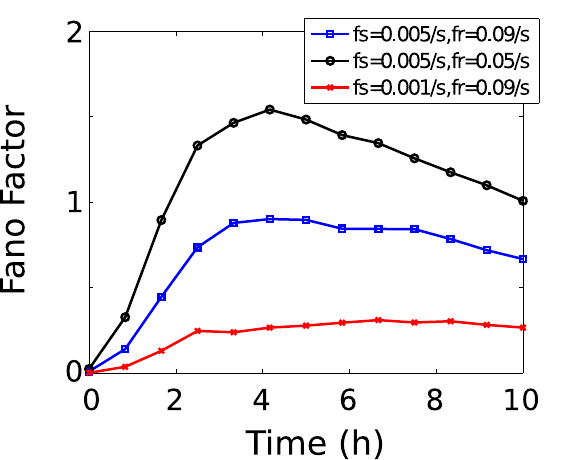} \label{fig:frfs-t} }
 	 \subfigure[]{    \includegraphics[width=0.5\textwidth]{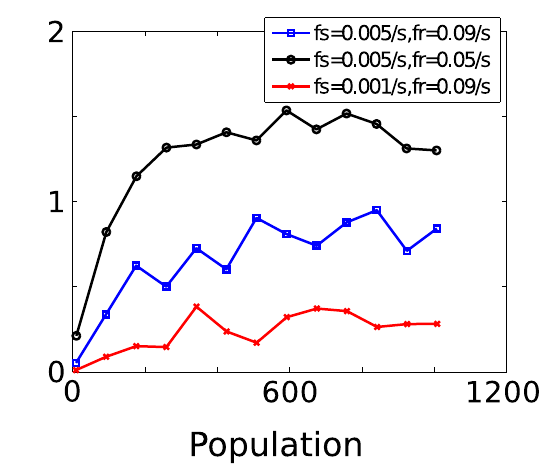} \label{fig:frfs-N} }
	  \subfigure[]{    \includegraphics[width=0.5\textwidth]{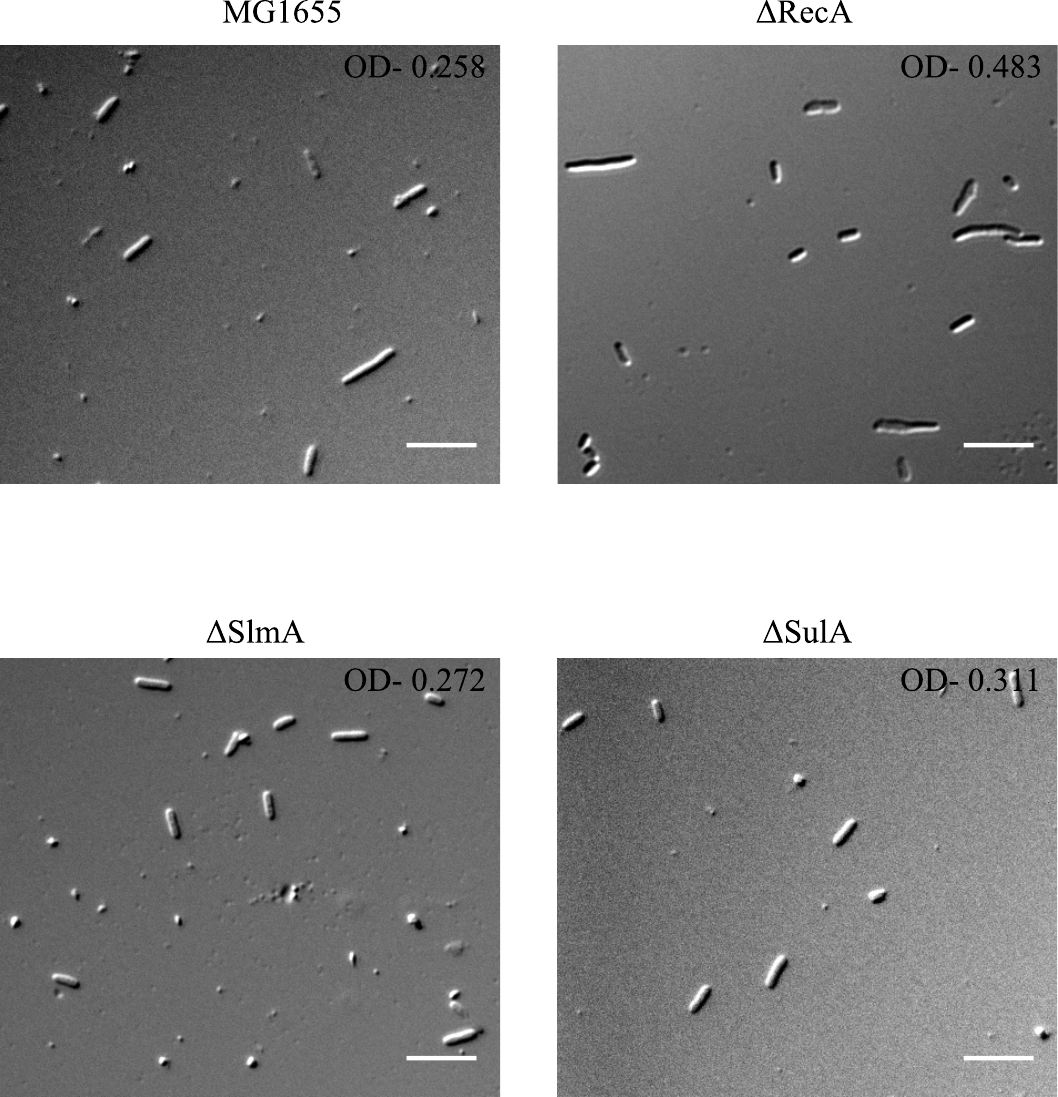} \label{fig:DIC} }
   	 \subfigure[]{    \includegraphics[width=0.5\textwidth]{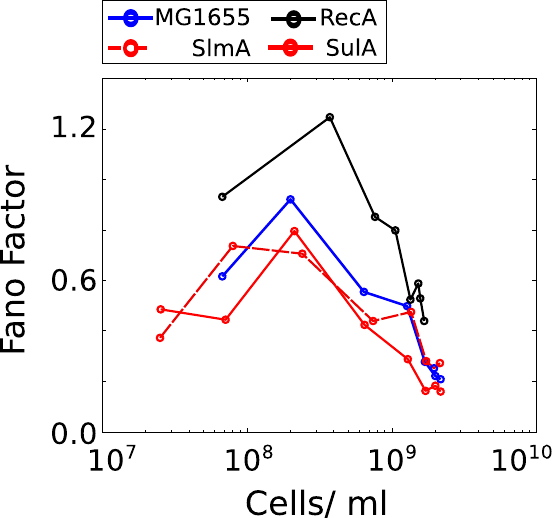} \label{fig:expFF} }\\

\caption{{\bf Effect of replication fork stochasticity.} \subref{fig:frfs-t},\subref{fig:frfs-t} The fano factor ($\sigma^2/\mu$) of cell length distributions is plotted for simulated populations with mean $\pm$s.d. for multiple runs (n=10) as a function of \subref{fig:frfs-t} time and \subref{fig:frfs-N} population size. Replication stochasticity of wild-type with $f_s=0.05$ $s^{-1}$ and $f_r=0.09$ $s^{-1}$ (blue) is compared to mutants with either $f_r= 0.05$ (black) or $f_s= 0.001$ (red). 
\subref{fig:DIC} Representative DIC microscopy images of wild-type (MG1655), $\Delta$recA, $\Delta$slmA and $\Delta$sulA from the same mid-log growth phase. Scale-bar 10 $\mu m$). \subref{fig:expFF} The fano factor of cell length is plotted as a function of cell density (cells/ml) for MG1655 (black), $\Delta$recA (red), $\Delta$sulA (green) and $\Delta$slmA (blue).}
\label{fig:exptsim}       
\end{figure}

\subsection{Replication fork stochasticity and cell size variability}
\label{sec:rfStoch}
To determine whether the model of replication-division coupling driven by stochasticity of RF dynamics could alone explain the trends in cell length data, we varied the stalling frequency ($f_s$) and recovery frequency ($f_r$), that determine the average state of each RF, based on a two-state model (stalled and recovered). The wild type RF stalling frequency ($f_r$) for {\it E. coli} MG1655 taken to be $0.005$ $s^{-1}$, while the value of $f_r$ was optimized to 0.09 $s^{-1}$, in order to achieve qualitative agreement of simulated cell length distributions with the experimental measures from {\it E. coli} wild-type cells grown at $37^oC$ in LB \cite{Gangan:2017aa}. We assume the value of the {\it in vivo} stalling frequency to be an order of magnitude smaller than the {\it in vitro} measurement of {\it E. coli} Pol-III holoenzyme $f_s = 0.02$ $s^{-1}$  \cite{Tanner2008}. In the absence of a measurement, this assumption is based on the expectation that inside the cell crowding and additional factors will enhance RF progression \cite{Nielsen2007,Pham2013}. The perturbations in simulations are either (a) decreased frequency of replication fork recovery ($f_r$) by two-fold, and (b) decrease of $f_s$ by five-fold. The (a) decreased $f_r$ is comparable to $\Delta recA$ mutant cells in which DNA RFs are more likely to stall as compared to wild-type (w.t.) \cite{Michel2007,Skarstad1988} due to the absence of RecA protein based recovery of stalls (Figure \ref{fig:recsulModel}). In contrast, (b) decreased $f_s$ is used to mimic
the absence of {\it SlmA}, which in its native form binds to DNA and prevents the division of cells when the DNA occludes the center of the cell ('nucleoid occlusion') \cite{Bernhardt2005} and the absence of {\it SulA}, which functions normally to inhibit cell division division when DNA damage (e.g. RF stalling) is detected by the SOS pathway by binding to FtsZ \cite{Huisman1984,Trusca:1998aa} as illustrated in Figure \ref{fig:recsulModel}. 

We find cell length variability (FF) of simulated strains with $f_r= 0.05$ $s^{-1}$ is higher, while that of strains with $f_s= 0.001$ $s^{-1}$ is lower than the `wild-type' throughout the growth period (Figure \ref{fig:frfs-t}), showing a saturation as a function of cell density (Figure \ref{fig:frfs-N}). Experimental evaluation of microscopic images of cells in wild-type ({\it E. coli} MG1655), $\Delta recA$ (lower $f_r$), $\Delta sulA$ (lower $f_s$) and $\Delta slmA$ (lower $f_s$)  seen before \cite{Athale2011,Gangan:2017aa} appears to suggest an increased variability in cell sizes of $\Delta recA$ cells, and a decreased variability of $\Delta sulA$ and $\Delta slmA$ cell (Figure \ref{fig:DIC}). This is quantitatively borne out by the quantification in terms of FF as a function of population density. 
Thus simulations we believe match experiments where $\Delta recA$ results in a higher frequency of replication stalling and leads to increased variability, while de-coupling the replication stalling and division machinery results in reduced variability in $\Delta sulA$ and $\Delta slmA$ strains. 

\section{Discussion}

We have developed a novel model of bacterial cell growth and division, which integrates sub-cellular processes with population dynamics. The agent-based multi-scale model integrates a stochastic, multi-fork chromosome replication model coupled to cell division, in the context of a growing population with a finite carrying capacity. Using this model, we examine the effect of growth-rate and -phase on cell size variability, to disambiguate previous experimental findings. The only source of stochasticity in the model is the probabilistic transition of DNA RFs between stalled and recovered states. Cells become filamentous, i.e. long, when they either do not complete DNA replication within a bacterial cell cycle or contain one or more stalled RFs at the time of division. The cell length growth rate ($g_L(t)$) on the other is assumed to be uniform for all cells in the population for a specific time point (t). The model predicts cell length variability is highest in the exponential growth phase as compared to the lag and stationary phases. The time of onset of increase in cell length variability from the lag to log phases differs between cultures simulated at different growth rates. However, when scaled by population size, the variability collapses onto the same curve across temperatures, suggesting instantaneous growth rate rather than growth phase or maximal growth rate of a batch culture is the most important factor determining replication stochasticity driven cell size variability. In order to test the predictive value of the model, we vary RF frequencies of stalling ($f_s$) and find cell size variability to change proportionately. The predictions are validated by comparison to the cell size variability of mutants- increased cell length variability of $\Delta recA$ with higher DNA RF stalling rates as compared to wild-type (w.t.) and decreased cell length variability of both mutants of $\Delta slmA$ and $\Delta sulA$. In its native form, $slmA$ binds to DNA and prevents the division of cells when the DNA occludes the center of the cell ('nucleoid occlusion'), that could occur due to stalled and incomplete replication. Additionally, {\it SulA} is triggered to inhibits cell division inhibition when {\it RecA} filaments assemble on stalled RFs. 

Due to the small number of parameters (Table \ref{tab:simparam}), this model is relatively tractable. Some parameters such as genome elongation rate, maximal doubling time 
are derived from experimental measurements in previously published work \cite{Gangan:2017aa} or measured in this study. Other parameters such as the frequencies of RF stalling and recovery are optimized to fit the extent of cell size distribution observed experimentally in wild-type cells. The $f_s$ of the {\it E. coli} Pol-III holoenzyme has been estimated using {\it in vitro} reconstitution to be 0.02 $1/s$ \cite{Tanner2008}. However, when such a high value was used in our model, it resulted in a far higher frequency of elongated cells, as compared to experiments, and as a result was not used. The lower frequency of stalling {\it in vivo} as compared to {\it in vitro} could be the result of the fact that {\it in vivo}, additional proteins such as RecA and RecBCD \cite{Michel2007}, helicases, single strand binding proteins and resolvases \cite{Cox2000}, which are all implicated in RF recovery pathways. This could also explain why cells of the $\Delta recA$ mutant are only slightly more variable than wild type cells (Figure \ref{fig:expFF}). In future {\it in vivo} single-molecule measurements that have been developed to follow replication proteins \cite{Uphoff2013}, combined with systematic screening of mutants, could be used to quantify RF dynamics more precisely.

The multi-fork replication model implemented here is based on the Cooper and Helmstetter (CH) model \cite{Cooper1968}, which predicts a change in $n_{RF}$ with changing growth rate. Our simulations demonstrate the early growth phase of cells contains multiple RFs (up to 8) and saturation phase cultures result in only a pair of RFs per cell. Importantly, the probability of replication stalling is modeled as an independent event, but cell division evaluates the total number of stalling events. As a result, cells experience a multiplicative effect of RF stochasticity on cell division. As a consequence, rapid growth results in higher cell size variability as compared to saturated growth. This clarifies the apparent contradictions in previous reports of temperature \cite{Trueba1982,Shehata1975} and nutrient dependence \cite{Kubitschek1983} of cell sizes, where it was unclear which factor primarily determined the distribution. We hypothesize that a simple explanation based on multiplicative probabilistic events, based on multi-fork replication underlie these effects. In future, experiments to count the RF numbers under these conditions could further test our hypothesis. 

Single-cell quantitative studies of bacterial growth have begun to reveal subtle effects of asymmetric division and aging in {\it E. coli} due to improvements in microscopy \cite{Wang2008}. A theoretical model of ``increment'' based control of cell size of {\it E. coli} and {\it Caulobacter crescentus} has been described \cite{Amir2014}. The variability of growth rates observed in single-cell continuous culture \cite{Wang2010} was explained by variations in the cell elongation rate ($g_L$) and cell size added \cite{Taheri-Araghi2014}, further confirming the ``increment'' model. While this model is consistent with experimental data, it lacks a molecular mechanism, that could cause such variation. Additionally in the absence of collective effects, the behavior only addresses single cell variability. Here in our work we have used a multi-scale approach to address both the morphological measurements seen in older work, as well as propose molecular processes that drive it.

Discrete models of bacterial populations have been widely used in the past to examine spatial patterns \cite{Ben-Jacob1994}, chemotaxis \cite{Emonet2005}, synthetic biology \cite{Gorochowski2012}, branching growth of colonies \cite{Farrell2013} and population effects in quorum sensing \cite{Mina2013}. However this is the first report of a replication-division coupling model that results in cell shape distribution as a function of population size and growth rate. 

Our model of replication stochasticity and the effect of multi-fork replication on cell size variability predicts population cell size variability, which is consistent with previous work. Our results also disambiguate some of the results obtained with respect to the effect of temperature, population density, time and growth rate. The simulations suggest that cell size variability in a population is most strongly determined by growth rate. Additionally we predict the effect on cell size variability of mutations in {\it $\Delta$recA} can be reproduced by an increased frequency of replication stalling, while {\it $\Delta$sulA} and {\it $\Delta$slmA} mutants will result in reduced effect of replication fork stalling on cell division. In conclusion this study could form the basis to use cell size variability to better understand the interactions between regulatory factors governing a phenotype as complex as cell size.


%

\section{Acknowledgements}
The work was funded by IISER core-funding and a Basic Biology Grant by the Dept. of Biotechnology, Govt. of India BT/PR1595/BRB/10/1043/2012.
We would like to thank Prasad Perlekar and Girish Deshpande for discussions. Hemangi Chaudhari was involved in early stages of the modeling work. Manasi S. Gangan performed the DIC microscopy of mutant {\it E. coli} strains and ran the image analysis code. 

\newpage
\bibliographystyle{elsarticle-num}
\bibliography{ecoli-CellGrowthModel}   
%
%

\end{document}